\documentclass[11pt,twoside]{article}
\usepackage{asp2004}
\usepackage{psfig}
\usepackage{epsf}
\usepackage{graphics}
\usepackage{lscape}

\begin{document}
\title{The Turn-off of Classical Novae Observed Through X-rays: Models versus ROSAT Observations of V1974 Cyg}
\author{Gl\`oria Sala and Margarita Hernanz}
\affil{Institut d'Estudis Espacials de Catalunya and Institut de Ci\`encies de l'Espai (CSIC). Campus UAB, Facultat de Ci\`encies, Torre C5-parell, 2n pis. E-08193 Bellaterra (Barcelona), Spain}

\begin{abstract}
Detection of X-ray emission from classical novae in their post-outburst stages
provides crucial information about the nova phenomenon. The soft X-ray emission
gives a direct insight into the remaining hot nuclear burning shell.
A numerical model for steady H-burning white dwarf envelopes has been developed
to study the post-outburst phases of classical novae. Properties of the X-ray
emission of the post-nova white dwarf are obtained,
and the results are compared to ROSAT observations of V1974 Cyg.
\end{abstract}

\section{White Dwarf Envelope Models with Steady H-Burning}

White dwarf envelope models with steady hydrogen burning have been
calculated for realistic compositions of white dwarf envelopes in
classical novae, and their evolution has been approximated as a
sequence of steady states. Three chemical compositions for ONe
novae have been considered, corresponding to models from
\cite{jh98} with 25\%, 50\% and 75\% mixing between the solar
accreted matter and the degenerate core. Results show that, for a
given white dwarf mass and composition, the envelope evolves in
the HR diagram increasing its effective temperature along a
plateau of quasi-constant luminosity, while its mass slowly
decreases due to hydrogen burning. When the envelope mass is
reduced to the minimum value for steady hydrogen burning, the
source turns-off. This occurs just when the maximum effective
temperature is reached.

Both the luminosity and the envelope mass during the plateau depend
on the white dwarf mass and hydrogen mass fraction. The behaviors of
luminosity and envelope mass as a function of white dwarf mass and
hydrogen mass fraction can be approximated by the expressions

\begin{center}
\(L_{\rm plateau}(L_{\odot })\simeq 5.95\times 10^{4}\left( M_{\rm
c}/M_{\odot} -0.536X_{\rm H}-0.14\right)\)

\(\log M^{\rm ONe}_{\rm plateau}(M_{\odot })\simeq 0.42X_{\rm
H}-\left( M_{\rm c}/M_{\odot}-0.13\right) ^{3}-5.26\)
\end{center}

\section{Comparison with ROSAT/PSPC Observations of V1974 Cyg}

\begin{figure}
\centering
\resizebox{13.0cm}{!}{\plottwo{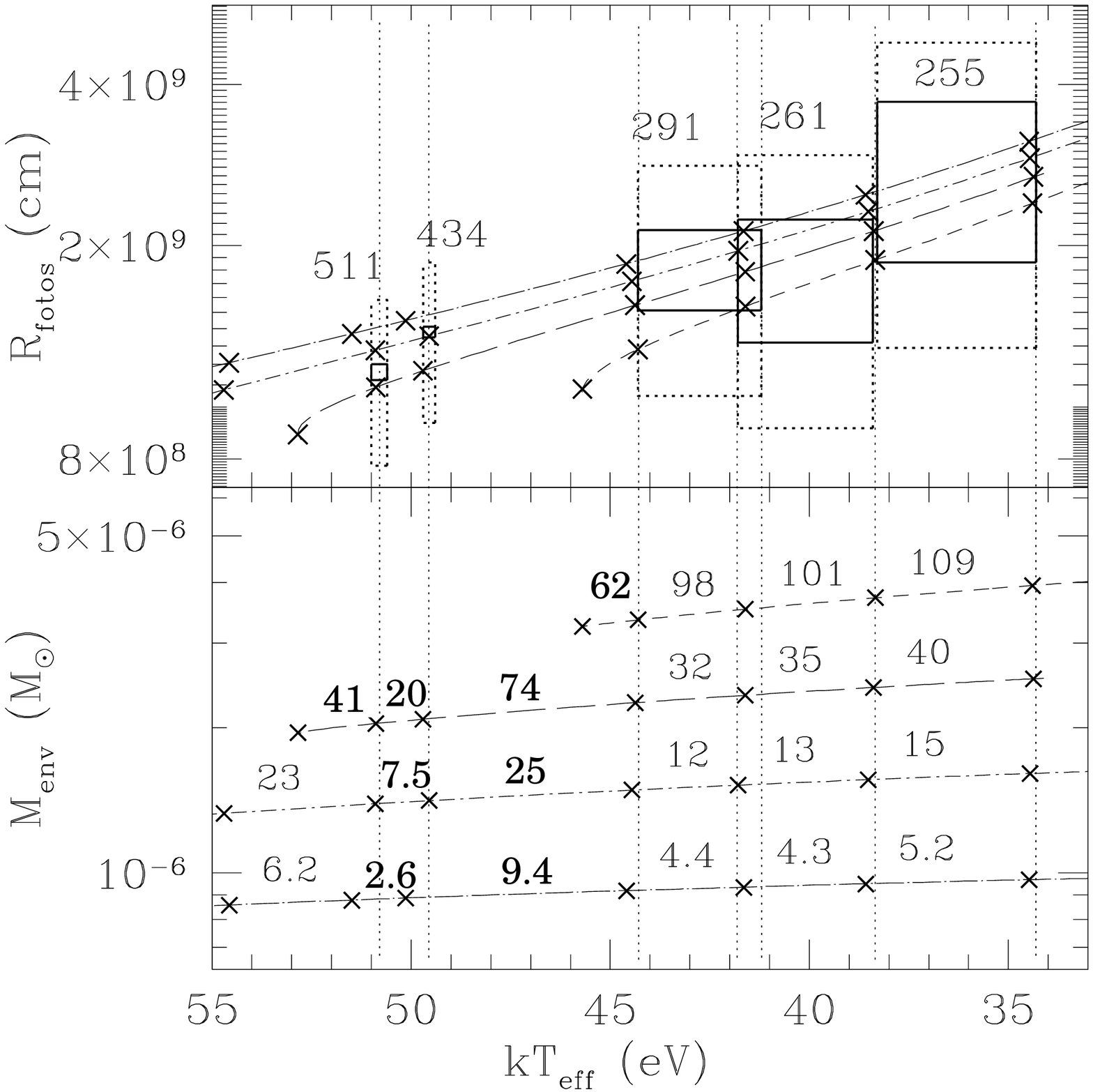}{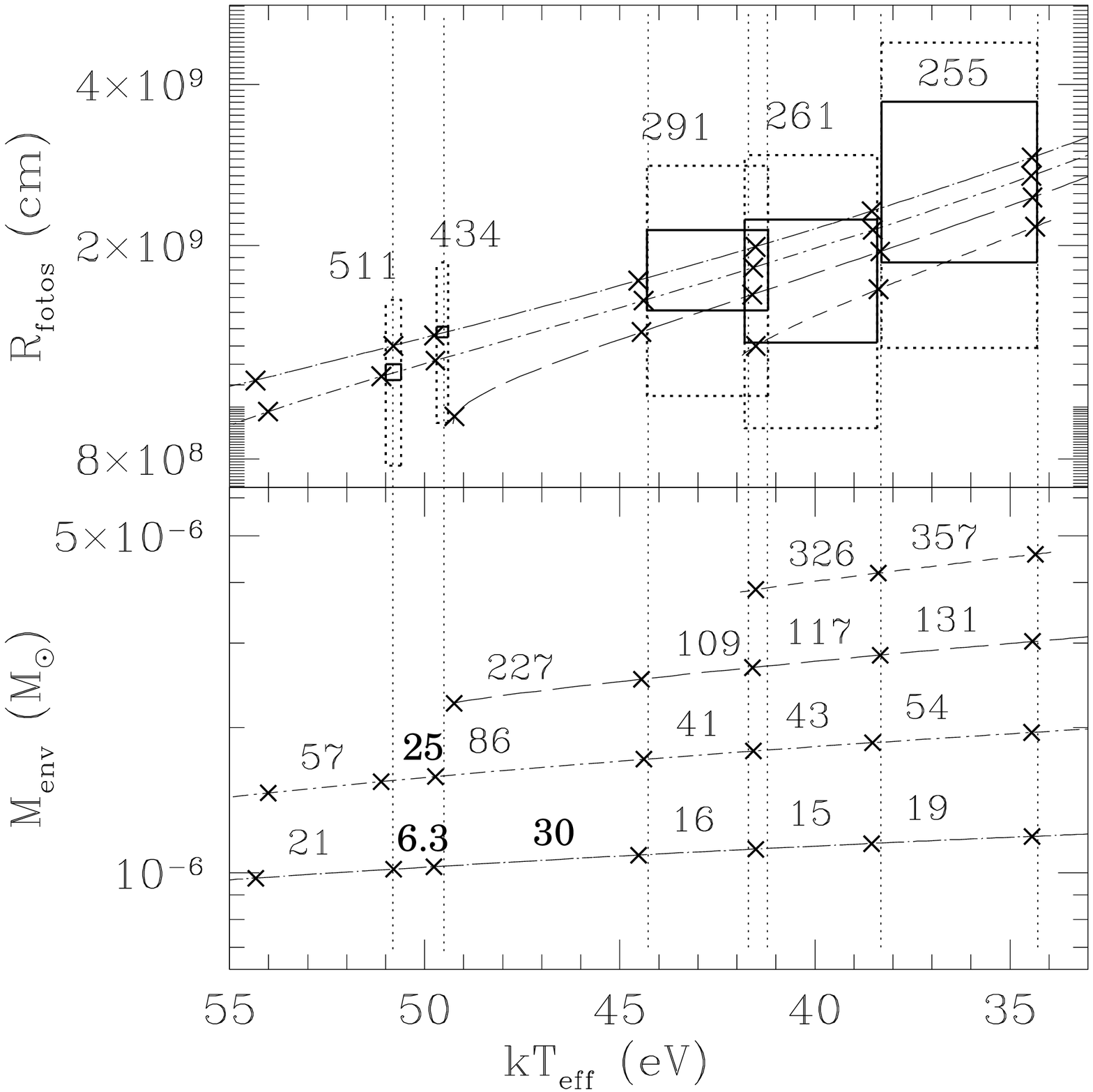}}
\caption{
Observational results for V1974 Cyg compared with our 
envelope models with 50\% (left) and 25\% (right)
mixing. \textbf{Upper panels:} 3$\sigma$ confidence contours
obtained by \cite{bal98} for V1974 Cyg 
on the R$_{\rm photos}$-kT$_{\rm eff}$ plane.
Results are shown for a distance of 2.5 kpc (solid contours)
and for a range of 1.7-3.2 kpc (dashed contours). 
Numbers indicate the day of ROSAT observation after
outburst. Models are plotted for 0.8 M$_{\odot}$
(short-dashed line), 0.9 M$_{\odot}$ (long-dashed), 1.0
M$_{\odot}$ (short dash-dot) and 1.1 M$_{\odot}$ (long dash-dot).
\textbf{Lower panels:} Envelope
masses for the models plotted in the upper panel. Numbers between
ticks indicate time in days needed for the envelope to evolve
due to pure hydrogen burning.}

\label{figures}
\end{figure}

We have compared  \cite{bal98} 
results on the soft component of
V1974 Cyg (Nova Cygni 1992) during the plateau phase with our
steady hydrogen burning white dwarf envelope models. In Fig.
\ref{figures}, the \( 3\sigma  \) contours of the spectral
parameters found by  Balman, Krautter, \& \"Ogelman (1998) are
plotted over the photospheric radius-effective temperature results
from our envelope models with 25\% and 50\% mixing. For each
model, the lower panels show the corresponding envelope mass with
the time intervals (in days) for quasi-static evolution.

It is clear that the large error bars on the photospheric radius,
due to the uncertainty on the distance, prevent any clear
comparison with the value predicted by the models. Nevertheless,
the comparison of the observed evolution of the effective
temperature alone with the models indicates that the soft X-ray
emission of V1974 Cyg is compatible with the evolution of an
envelope with steady hydrogen burning, on either a 0.9 M$_{\odot}$ 
white dwarf with 50\% mixing, or 1.0 M$_{\odot}$ 
with 25\% mixing. Note that this comparison based on
the effective temperature is independent from the distance
determination.

In any case, the envelope mass for stable hydrogen burning ($\sim
10^{-6}$ M$_{\odot}$) is much smaller than the mass needed to
trigger the nova outburst for these masses and compositions ($\sim
10^{-5}$ M$_{\odot}$, Jos\'e \& Hernanz 1998). As the ejected
fraction, according to hydrodynamical models, is not sufficient to
reduce the envelope mass down to the values for stable hydrogen
burning, it is clear that some extra mass loss mechanism must be
present after the outburst.

\end{document}